\documentclass[prd,nofootinbib,preprint,superscriptaddress,letter]{revtex4}

\usepackage{amsfonts}
\usepackage{amsmath}
\usepackage{amsthm}
\usepackage{amssymb}
\usepackage{array}
\usepackage{dcolumn}
\usepackage[svgnames]{xcolor}
\usepackage[hidelinks]{hyperref}
\hypersetup{
    colorlinks=true,
    linkcolor=NavyBlue,
    filecolor=NavyBlue,
    urlcolor=NavyBlue,
    citecolor=NavyBlue,
    }

\usepackage{graphics}
\usepackage{tikz}
\usepackage{graphicx}
\usetikzlibrary{positioning}
\usepackage{adjustbox}
\usepackage{breqn}
\usepackage{braket}
\usepackage{enumitem}

\usepackage{tikz-feynman,contour}
\tikzfeynmanset{compat=1.1.0}
\usepackage{tikzsymbols}

\def\be{\begin{equation}}
\def\ee{\end{equation}}
\def\bea{\begin{eqnarray}}
\def\eea{\end{eqnarray}}

\def\gev{\, {\rm GeV}}

\newcommand{\gsim}{\lower.7ex\hbox{$\;\stackrel{\textstyle>}{\sim}\;$}}
\newcommand{\lsim}{\lower.7ex\hbox{$\;\stackrel{\textstyle<}{\sim}\;$}}

\newcommand{\dd}{\text{d}}

\begin{document}

\title{Gravitational Wave Production and Baryogenesis in a Simple Left-Right model }

\author{Arnab Dasgupta}
\email{arnabdasgupta@pitt.edu}
\affiliation{Pittsburgh Particle Physics$,$ Astrophysics$,$ and Cosmology Center$,$
Department of Physics and Astronomy$,$ University of Pittsburgh$,$ Pittsburgh$,$ PA 15206$,$ USA}

\author{Matthew Knauss}
\email{mhknauss@wm.edu}
\affiliation{High Energy Theory Group$,$\ Department of Physics$,$ William\ \&\ Mary$,$ Williamsburg$,$ VA 23187$,$ USA}

\author{Marc Sher}
\email{mtsher@wm.edu}
\affiliation{High Energy Theory Group$,$\ Department of Physics$,$ William\ \&\ Mary$,$ Williamsburg$,$ VA 23187$,$ USA}

\begin{abstract}
The left-right model with the simplest Higgs structure has a $SU(2)_L$ doublet $H_L$, a $SU(2)_R$ doublet $H_R$, and a singlet $\sigma$ that couples to the doublets as $\sigma(|H_L|^2-|H_R|^2)$.   The left-right symmetry has $H_L\leftrightarrow H_R$ and $\sigma\leftrightarrow-\sigma$.   We study gravitational wave production and baryogenesis in the electroweak phase transitions in the model.  For two benchmark points, $\sigma$ first gets a vev, followed by $H_R$ and then $H_L$.   An interesting feature is that the vev of $H_L$ is initially much higher than its zero temperature value, leading to a more strongly first-order transition and higher frequency gravitational waves.    An unusual  benchmark point has the $\sigma$ vev at zero when the electroweak symmetry breaks.  This results in both $H_L$ and $H_R$ vevs being equal and in the multi-TeV range.   At a lower temperature, the $\sigma$ vev turns on, breaking the left-right symmetry and $H_L$ drops to its standard model value.    Since the electroweak symmetry is broken at the multi-TeV scale, the frequency of gravitational waves will be much higher than usual.   Baryogenesis is also discussed.
\end{abstract}

\maketitle

\section{Introduction}
One of the mysteries of cosmology is the matter-antimatter asymmetry.     The current value of the asymmetry is $n_B/n_\gamma \simeq 6 \times 10^{-10}$ \cite{Planck:2018vyg}.  Since baryon number violation is required to generate such an asymmetry, initial discussions of baryogenesis focused on grand unified theories.   However,  it was recognized that electroweak sphalerons could wipe out an initial baryon asymmetry, which led to interest in baryogenesis on the electroweak scale, where sphalerons can generate the asymmetry \cite{Fukugita:1986hr,kuzmin,Cohen:1993nk,Rubakov:1996vz,Trodden:1998ym,Morrissey:2012db}.   The out-of-equilibrium Sakharov condition\cite{Sakharov} is most easily satisfied via a first-order phase transition.    

 The Standard Model does not have a first-order phase transition\cite{Rummukainen:1998as} and does not have sufficient CP violation.   The simplest and most well studied extension is the Two-Higgs Doublet Model (2HDM) \cite{Branco:2011iw}.   This can have a first-order phase transition and there is substantial literature on baryogenesis in 2HDMs, in the CP conserving case\cite{Bochkarev:1990fx,Turok:1990zg,Turok:1991uc,Funakubo:1993jg,Davies:1994id,Funakubo:1995kw,Funakubo:1996iw,Cline:1995dg,Fuyuto:2015jha,Chiang:2016vgf,Dorsch:2013wja,Dorsch:2014qja,Andersen:2017ika,Bernon:2017jgv,Kainulainen:2019kyp,Bittar:2025lcr}, see Ref. \cite{Basler:2016obg} for an extensive review.  In addition, studies of electroweak baryogenesis have been carried out in the 2HDM with explicit CP violation\cite{Cline:1996mga,Fromme:2006cm,Cline:2011mm,Dorsch:2016nrg,Haarr:2016qzq,Basler:2017uxn,Wang:2019pet,Enomoto:2021dkl}, see Ref. \cite{Basler:2021kgq} for a recent review.

Are there any signatures of electroweak baryogenesis?   Of course, many models do have collider signatures that can be tested at the LHC.    With the recent discovery of gravitational waves, however, the possibility of gravitational wave signatures from bubble wall collisions becomes attractive\cite{Apreda:2001us,Grojean:2006bp,Leitao:2012tx,Hindmarsh:2013xza,Kakizaki:2015wua,Chala:2018opy,Morais:2019fnm,Caprini:2019egz,Goncalves:2021egx,Baldes:2021vyz,Azatov:2021irb,Benincasa:2022elt,Huang:2022vkf,Dasgupta:2022isg,Cao:2022ocg,Chatterjee:2022pxf,Ghosh:2022fzp}.  The frequency range of these signatures\cite{Caprini:2019egz}  is in the range of LISA, BBO or DECIGO and might be detectable.

Generally a baryon asymmetry can best be produced from a slow moving bubble wall, whereas gravitational wave production best requires a rapidly moving bubble wall.  While some successful models do exist\cite{Huang:2016odd,Beniwal:2017eik,Demidov:2017lzf,Goncalves:2023svb}, they generally require complicated Higgs sectors and in 2HDM models may be incompatible with electron EDM measurements (although some regions of parameter space may in some cases be allowed)\cite{Dorsch:2016nrg,Zhou:2020irf,Ellis:2022lft,Lewicki:2021pgr}.    This led to an interest in multi-step electroweak phase transitions in which baryogenesis and gravitational wave production can be generated from different transitions.   

However, Cline and Kainulainen\cite{Cline:2020jre} recently have shown that a fast moving wall can still generate a sufficient baryon asymmetry\footnote{A precise determination of the bubble wall velocity is crucial.   Very recently, Carena et al.\cite{Carena:2025flp} have studied this in great detail. We will use the method of Cline and Kainulainen, but the reader is referred to Carena et al. for a more precise calculation.}. See Ref. \cite{Croon:2023zay} for a detailed discussion and numerous references.   As a result, it is possible that both the baryon asymmetry and gravitational wave production could occur at the same transition although this will not occur in the Standard Model or typical 2HDMs.    We will therefore not require that baryogenesis and gravitational wave production occur at different transitions.

Some of the earliest extensions of the Standard Model are left-right symmetric models\cite{Pati:1974yy,Mohapatra:1974hk,Mohapatra:1974gc,Senjanovic:1975rk}.   Here, one step can break the left-right symmetry and the other can break the electroweak gauge symmetry.   These models are based on the gauge group $SU(2)_L\times SU(2)_R \times U(1)$ and have received a great deal of attention. They differ in having a variety of fermion and scalar sectors.   The most common scenario has two $SU(2)$ triplet Higgs with one breaking the $SU(2)_R$ symmetry and also has a bidoublet field with includes the Standard Model Higgs and provides masses for the fermions.   This is a fairly complicated scalar sector that has doubly charged scalars and many other states.   The phase transition in this model was studied long ago \cite{Barenboim:1998ib} and the production of gravitational waves has also been examined and baryogenesis briefly discussed \cite{Brdar:2019fur,Li:2020eun, Wang:2024wcs}.    The simplest possible Higgs sector involves just two doublets, with fermion masses generated by a see-saw mechanism.   Gravitational wave production in this model was studied in Ref. \cite{Graf:2021xku}, but the model in that case is not left-right symmetric (since the $SU(2)_L$ and $SU(2)_R$ gauge couplings are different); one can impose left-right symmetry in the model at the cost of non-renormalizable terms.   Another scalar sector has  two doublets plus a bidoublet\cite{Senjanovic:1978ev,Mohapatra:1977be}.  Gravitational wave production in this model was recently studied in Ref. \cite{Karmakar:2023ixo}.    Another model has four doublets\cite{Ma:1989tz,Borah:2017leo}.

In this paper, we will focus on a left-right symmetric model without scalar triplets or bi-doublets.   The model was discussed by Gu\cite{Gu:2010yf} and contains two Higgs doublets, a left-handed doublet $H_L$, and right-handed doublet $H_R$ and a singlet $\sigma$.  The left-right symmetry requires $H_L \leftrightarrow H_R$ and $\sigma \leftrightarrow -\sigma$.    It is a multi-step transition in which $\sigma$ gets a vev breaking the left-right symmetry (since it is a singlet, this vev doesn't break the electroweak symmetry), and the electroweak symmetry is broken when $H_L$ or $H_R$ (or both) get a vev.  Fermion masses are generated by integrating out heavy fermions. Although the definition of ``minimal left-right symmetric model" is subjective, we consider this model to have the minimal scalar sector - it is the only such sector with no charged or pseudoscalar Higgs bosons.

Gu studied the baryon asymmetry in the model.  Here we will focus on both the baryon asymmetry and gravitational wave production.   As we will see, the model has a very unusual history, in that the phase transition breaking the electroweak symmetry can occur while the model is still left-right symmetric, leading to $H_L$ and $H_R$ both getting vev at a fairly high temperature.   In Section II, the model is presented with the various stages of symmetry breaking.   The symmetry-breaking pattern is discussed in Section III.  Gravitational wave production is discussed in Section IV and baryogenesis in Section V. Section VI contains our conclusions.

\section{Model}
The particle content for this left-right $SU(3)_c \times SU(2)_L \times SU(2)_R \times U(1)_{B-L}$ model\cite{Gu:2010yf} is given in Table 1.
\begin{table}[!h]
    \centering
    \begin{tabular}{|c|c|c|}
       \hline \hline   
         & $n_f$ & $SU(3)_c \times SU(2)_L \times SU(2)_R \times U(1)_{B-L}$  \\
       \hline
        $q_L = \begin{pmatrix} u \\ d\end{pmatrix}$ & 3 & (3,2,1,1/3) \\
        $q_R = \begin{pmatrix} \overline{u} \\ \overline{d} \end{pmatrix}$ & 3 & ($\bar{3}$,1,2,1/3) \\
        $\ell_L = \begin{pmatrix} \nu \\ e \end{pmatrix}$ & 3 & (1,2,1,-1) \\
        $\ell_R = \begin{pmatrix} \overline{\nu} \\ \overline{e} \end{pmatrix}$ & 3 & (1,1,2,1) \\
        $U_L$  & 3 & (3,1,1,1/3) \\
        $U_R$  & 3 & ($\bar{3}$,1,1,1/3) \\
        $D_L$  & 3 & (3,1,1,1/3) \\
        $D_R$  & 3 & ($\bar{3}$,1,1,1/3) \\
        $E_L$  & 3 & (1,1,1,-1) \\
        $E_R$  & 3 & (1,1,1,-1) \\
        $N_L$  & 3 & (1,1,1,-1) \\
        $N_R$  & 3 & (1,1,1,-1) \\
        $H_{L}$  & 1 & (1,2,1,0) \\
        $H_{R}$  & 1 & (1,1,2,0) \\
        $\sigma$ & 1 & (1,1,1,0) \\
        \hline
    \end{tabular}
    \caption{In the above table we tabulate the particle content for the Left-Right model with a minimal scalar sector.}
    \label{tab:content}
\end{table}

The scalar potential is given by:
\begin{align}
V_0 &= -\mu^2_1 \sigma^2 -\mu^2_2 (H^{\dagger}_L H_L + H^{\dagger}_R H_R) + \mu_3 \sigma \left(H^\dagger_L H_L - H^\dagger_R H_R \right)+\lambda_\sigma\sigma^4 \nonumber \\
    &+ \lambda_H \left[ (H^\dagger_L H_L)^2 + (H^\dagger_R H_R)^2\right] + \lambda_{LR} H^\dagger_L H_L H^\dagger_R H_R + \lambda_{\sigma H} \sigma^2 \left(H^\dagger_L H_L + H^\dagger_R H_R\right),
\end{align}
and the Yukawa Lagrangian is  \begin{align}
    \mathcal{L}_Y &= -y_D(\bar{q}_L H_L D_R +\bar{q}_RH_RD_L) - M_D \bar{D}_L D_R\\
    & -y_U(\bar{q}_L \tilde{H}_L U_R +\bar{q}_R\tilde{H}_RU_L) - M_U \bar{U}_L U_R\\
     & -y_E(\bar{\ell}_L H_L E_R +\bar{\ell}_RH_RE_L) - M_E \bar{E}_L E_R
     \\
     & -y_N(\bar{\ell}_L \tilde{H}_L E_R +\bar{\ell}_R \tilde{H}_R E_L) - M_N \bar{N}_L N_R
\end{align}
Integrating out the heavy fermions, the masses of the SM fermions are as follows:
\begin{align}
    m_u = y_U\frac{v_Lv_R}{M_U}y^\dagger_U \\
    m_d = y_D\frac{v_Lv_R}{M_D}y^\dagger_D \\
    m_l = y_E\frac{v_Lv_R}{M_E}y^\dagger_E \\
    m_\nu = y_N\frac{v_Lv_R}{M_N}y^\dagger_N 
\end{align}

\subsection{Bounded From Below}

One can constrain the parameters by requiring that the vacuum is stable at large field values.    These constraints then apply to the quartic parameters.  We have two doublets and a singlet.   The bounds on the neutral components in a three-Higgs model are given explicitly in \cite{Klimenko:1984qx,Kannike:2012pe}.     It has been noted \cite{Boto:2022uwv} that there is no simple expression for bounded-from-below constraints including both the charged and neutral directions; however, our model contains no charged or pseudoscalar Higgs bosons since they are eaten by the left- and right-handed gauge bosons.    Thus, we focus on the neutral components and the resulting bounds are necessary and sufficient.   The bounds are

\begin{align}
\lambda_\sigma \geq 0 \\
\lambda_H \geq 0 \\
\lambda_H + \frac{1}{2}\lambda_{LR} \geq 0 \\
2\sqrt{\lambda_\sigma\lambda_H} + \lambda_{\sigma H} \geq 0
\end{align}
\be
(\lambda_H+\frac{1}{2}\lambda_{LR})\sqrt{2\lambda_\sigma} + \lambda_{\sigma H}\sqrt{2\lambda_H} + (2\sqrt{\lambda_\sigma\lambda_H} + \lambda_{\sigma H}) \sqrt{(\lambda_H+\frac{1}{2}\lambda_{LR})} \geq 0
\ee

and we will restrict our study to parameters that satisfy these conditions. 

\subsection{Scalar masses}

One can eliminate the $
\mu^2$ parameters in terms of the the vevs of $H_L,H_R$ and $\sigma$:
\begin{align}
    \mu^2_2 &= \frac{1}{4}\left(\lambda_{LR} +2 \lambda_H)(v^2_L + v^2_R\right) + \lambda_{\sigma H} v^2_\sigma \nonumber \\
    \mu^2_1 &= \frac{1}{2}\lambda_{\sigma H}(v^2_L + v^2_R) + \frac{\left(\lambda_{LR} -2 \lambda_H)(v^2_L - v^2_R\right)^2}{16v^2_\sigma} + 2\lambda_\sigma v^2_\sigma \nonumber \\
    \mu_3 &= \frac{\left(\lambda_{LR} -2 \lambda_H)(v^2_L - v^2_R\right)}{4v_\sigma}
\end{align}

Then we compute the scalar mass matrix, defining $\lambda_v\equiv \frac{(\lambda_{LR}-2\lambda_H)(v_L^2-v_R^2)}{4v_\sigma}$,

\begin{align}
    m_h^2 &= 2\begin{pmatrix}
        2\lambda_H v^2_L & \lambda_{LR} v_L v_R & 2\lambda_{\sigma H} v_L v_\sigma + \lambda_v v_L\\
        \lambda_{LR} v_L v_R & 2 \lambda_H v^2_R & 2\lambda_{\sigma H} v_R v_\sigma - \lambda_v v_R\\
        2\lambda_{\sigma H} v_L v_\sigma + \lambda_v v_L& 2\lambda_{\sigma H} v_R v_\sigma - \lambda_vv_R & 8\lambda_\sigma v^2_\sigma - \lambda_v\frac{(v^2_L - v^2_R)}{2v_\sigma}
    \end{pmatrix} \nonumber \\
    &= \mathcal{O}.m_D.\mathcal{O}^T \\
    m_D &= \begin{pmatrix}
        m^2_{Higgs} & 0 & 0\\
        0 & m^2_{H_R} & 0 \\
        0 & 0 & m^2_\sigma 
    \end{pmatrix}; \quad \mathcal{O} = \begin{pmatrix}
        c_{12}c_{13} & - c_{13}s_{12} & s_{13} \\
        c_{23}s_{12}-c_{12}s_{13}s_{23} & c_{12}c_{23}+s_{12}s_{13}s_{23} & c_{13}s_{23} \\
        -c_{12}c_{23}s_{13} - s_{12}s_{23} & c_{23}s_{12}s_{13} - c_{12}s_{23} & c_{13}c_{23}
    \end{pmatrix}
\end{align}

Of course, one must ensure that mixing between the $125$ GeV Higgs and the other neutral scalars is small enough to satisfy LHC bounds.   This will generally be the case since $v_L << v_R$.
\subsection{Phenomenological bounds}

A significant constraint on many multi-Higgs models arises from the $S$ and $T$ parameters.   These do not provide a major constraint here because we have no charged scalars or pseudoscalars.   

However, there are bounds on right-handed W masses.   Langacker and Sankar provided a detailed discussion of the bounds under a wide variety of assumptions\cite{Langacker:1989xa}.    We will simply use the strongest bound\footnote{In many models, detailed LHC analyses will provide stronger bounds.   We can accommodate these bounds, although the allowed parameter-space will shrink} in the Particle Data Group\cite{Workman:2022ynf} listing of $715$ GeV\cite{Czakon:1999ga}.   This implies (since the gauge couplings are very similar) $v_R > 8.8 v_L$.    In the next two sections, we will use a benchmark point with a smaller $W_R$ mass for illustrative purposes, but one must keep in mind that only masses in excess of $715$  GeV are phenomenologically acceptable.

\section{Phase Transitions and Breaking pattern}

In order to explore the transitions in detail, let us first focus on the effective potential. The complete effective potential consists of the tree-level contribution, the Colemen-Weinberg (CW) one-loop part, and then the corrections coming from the background temperature, which are as follows:

\begin{align}
    V_{effective} &= V_0(v_L,v_R,v_\sigma) + V_{CW}(v_L,v_R,v_\sigma) + V_{FT}(v_L,v_R,v_\sigma,T) + V_{daisy}(v_L,v_R,v_\sigma,T)
\end{align}

The Coleman-Weinberg (CW) part is stemming from the complete one-loop effective potential which is in general is written as:
\begin{align}
    V_{CW}(v_L,v_R,v_\sigma) &= \frac{1}{64\pi^2}\left[Tr\left[m^4_D\left(\log\frac{m^2_D}{\mu^2} - \frac{3}{2}\right)\right] + 6\sum_{i=L,R}m^4_{W_i}\left(\log\frac{m^2_{W_i}}{\mu^2} - \frac{5}{6}\right) \right. \nonumber \\
    &+ \left. 3\sum_{i=L,R}m^4_{Z_i}\left(\log\frac{m^2_{Z_i}}{\mu^2} - \frac{5}{6}\right) - 6\sum_{f=u,d,l,\nu}m^4_{f}\left(\log\frac{m^2_{f}}{\mu^2} - \frac{3}{2}\right)\right].
\end{align}
The temperature dependent part is given as:
\begin{align}
    V_{FT} &= \frac{T^4}{2\pi^2} \left[J_B\left(\frac{Tr[m^2_D]}{T^2}\right) + 6\sum_{i=L,R}J_B\left(\frac{m^2_{W_i}}{T^2}\right) + 3\sum_{i=L,R}J_B\left(\frac{m^2_{Z_i}}{T^2}\right) - 6\sum_{f=u,d,l,\nu}J_F\left(\frac{m^2_{f}}{T^2}\right) \right],
\end{align}
where the thermal functions $J_{B/F}$ are the integrals for bosons/fermions given as:
\begin{align}
    J_{B/F}(y^2) &= \int^\infty_0 dx\ x^2 \log\left(1 \mp e^{-\sqrt{x^2 + y^2}} \right).
\end{align}
The $\mu$ in the above equation is given as $\mu = \sqrt{v^2_L + v^2_R + v^2_\sigma}$ and the last expression $V_{daisy}$ comes from the daisy diagrams
\begin{align}
    V_{daisy} &= -\frac{T}{12\pi}\sum_i\left[(m^2_i + \Pi^2_i)^{3/2} - m_i^2\right],
\end{align}
where $\Pi_i$ is the thermal self energy correction to the field dependent masses.

\begin{table}[]
\centering
\begin{tabular}{|c|c|c|c|}
\hline\hline
 & BP1 & BP2 & BP3\\
 \hline
 $\lambda_\sigma$ & 0.065 & 0.014 & 0.0047\\
 $\lambda_H$ & 8 & 2.56 & 2.32\\
 $\lambda_{LR}$& -6.73&-2.11 & -2.12\\
 $\lambda_{\sigma H}$& 0.19&0.045 & 0.012\\
 $m_H$ (TeV) & 3.1 & 4 & 5\\
 $m_S$ (TeV) & 6.2 & 2 & 2.5\\
 $m_T$ (TeV) & 1.5 & 2 & 3.2\\
 $m_{W_R}$ (GeV) & 364 & 813 & 1070\\
 $v_S$ (TeV) & 12 & 8 & 18\\
 $v_R$ (TeV) & 1.11 & 2.5  & 3.3\\
 $T_{LR}$ (TeV) & 1.2 & & \\
 $T_L$ (GeV) &  & 485 & 668\\
 $T_R$ (TeV) &   & 1.9 & 2.4\\
 \hline
 \end{tabular}
 \caption{ Three representative benchmark points which satisfy the constraints of Section II except for the $W_R$ mass for the first benchmark point. \ {The vevs listed are the zero-temperature vevs}}
 \label{tab:benchmark_points}
\end{table}

The scalar potential contains seven free parameters.   These become the zero-temperature vevs $v_L, v_R$ and the observed Higgs mass, giving four free parameters (plus $v_R$, which will give the $W_R$ mass).     We will choose parameters such that the singlet vev is around $10$ TeV.   A set of benchmark points that satisfies all the bounds of the last section (except in one case, as noted earlier, the $W_R$ mass) as well as the temperatures and vevs for each of these points is given in Table \ref{tab:benchmark_points}.

\begin{figure}
    \centering
    \includegraphics[width=0.6\linewidth]{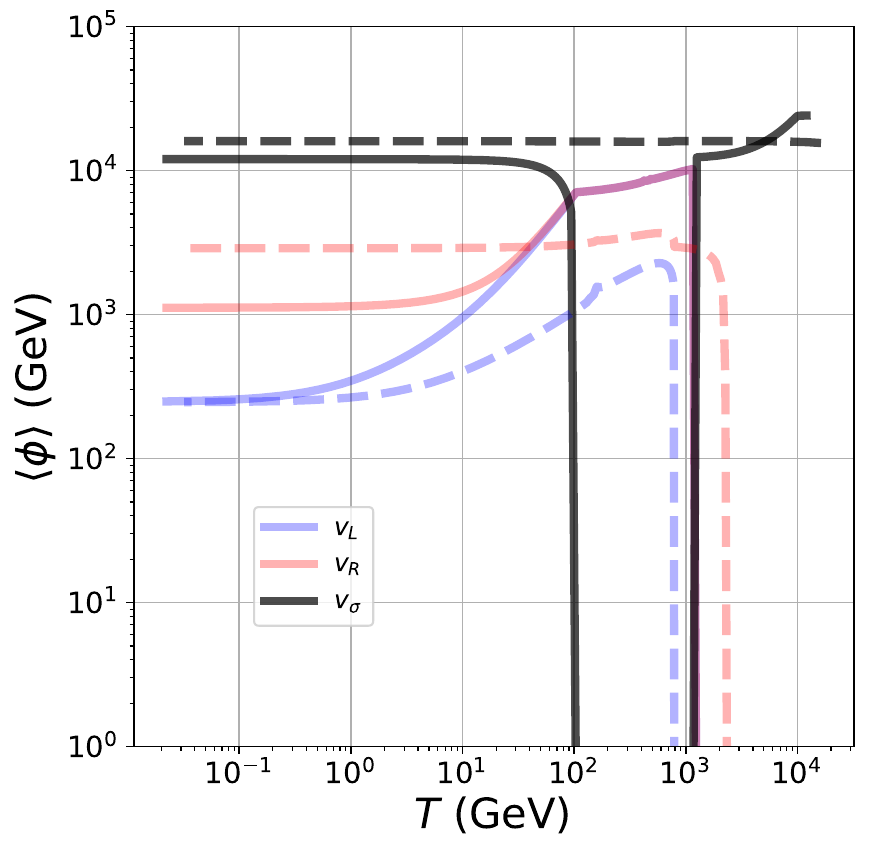}
\caption{ The vevs as a function of temperature for two benchmark points.   The solid lines correspond to benchmark point 1 and the dashed line to benchmark point 2. Benchmark point 3 is very similar to benchmark point 2. {At extremely high temperatures, the vev of the $\sigma$ is also zero in all cases considered.}} 
    \label{fig:1}
    \end{figure}

{Primarily there are two paths for the vevs from the fully symmetric phase $(0,0,0)$ to fully broken phase $(v_L,v_R,v_\sigma)$. The first path is when the $\sigma$ gets a vev $v_\sigma$ and then gradually restores back whilst $v_L=v_R$ gets a vev. And then later the $\sigma$ again gets a vev leading to the $L-R$ breaking. The second path, for which the mass of the $W_R$ is not constrained, is when the $\sigma$ gets a vev (break the $L-R$ symmetry) then the $H_R$ gets and vev and then finally $H_L$ gets a vev. }

 The temperature dependences of the vevs are shown in Figure 1 for two particular benchmark points.     For benchmark point 2 (the dashed line), the $\sigma$ vev is non-zero at a very high temperature - this breaks the left-right symmetry and the $H_R$ vev turns on next and the $H_L$ vev last.  It is interesting and unusual that the latter vev goes to a fairly high value before relaxing to its zero temperature value.    We have not shown the curve for benchmark point 3 since it is very similar to that of benchmark point 2.     {Although not shown explicitly on the plot, the vev of sigma does drop to zero a higher temperatures.}

 The more interesting benchmark point is benchmark point 1 (the solid line).  Here, the $\sigma$ turns on first.  The left-right symmetry is broken but both vevs of $H_L$ and $H_R$ are zero due to the high temperature.   At a lower temperature, $v_\sigma$ drops to zero, the left-right symmetry is restored, and ${\bf both}$ $H_L$ and $H_R$ get identical vevs, in the multi-TeV range.   Then the $\sigma$ vev returns, breaking the left-right symmetry and the vev of $H_L$ drops to its zero temperature value.   Although interesting, benchmark point 1 is phenomenologically unacceptable due to the light $W_R$ mass.

Note that $v_L/T_c$ for the transition where the vev of $H_L$ turns on for benchmark points 2 and 3 is much larger than 1, and for benchmark point 1 is also substantially larger than 1 thus the transition is strongly first-order and can give both baryogenesis and gravitational wave production.

 Our point in this paper is not to rely on a specific model but to show that models exist in which the electroweak transition temperature is much higher than usual -- in the multi-TeV range.\footnote{After this calculation was completed, a more consistent resummation technique was developed by Bittar, Roy and Wagner\cite{Bittar:2025lcr} - this might change the specific results in this model but won't change the basic result.}  That is the case for all three benchmark points.  The most interesting pattern of benchmark point 1 is not phenomenologically acceptable, but in more complicated models this pattern could occur.

\section{Gravitational Wave production}
The nucleation rate per unit volume  is 
given by~\cite{Linde:1977mm,Linde:1981zj} 
\begin{equation}
    \Gamma(T) = [A(T)]^4\exp[-S(T)] \, ,
    \label{eq:bubble}
\end{equation}
where $A$ is a pre-factor of mass dimension one (see below), and $S$ is the Euclidean bounce action. At zero temperature, the configuration minimizing the action is $O(4)$-symmetric, and $S\equiv S_4$, where 
\begin{equation}
 S_4 = 2\pi^2 \int_0^\infty {\rm d} r\: r^3 \left[\frac{1}{2}\left(\frac{{\rm d}\phi}{{\rm d} r}\right)^2+V_{\rm eff}(\phi,0)\right] \, ,
\label{eq:S4}   
\end{equation}
and can be estimated using the saddle-point approximation from the equation of motion:
\begin{equation}
    \frac{{\rm d}^2 \phi}{{\rm d} r^2} + \frac{3}{r} \frac{{\rm d}\phi}{{\rm d}r} - \frac{\partial V_{\rm eff}}{\partial \phi}
= 0 \, ,
\end{equation}
with the boundary conditions
\begin{equation}
    \frac{\dd\phi}{\dd r}(r=0)=0 \, , \qquad \phi(r=\infty)=0 \, .
    \label{eq:boundary}
\end{equation}
In the above equation, the first condition is for the solution to be regular at the center of the bubble, and the second one is to describe the initial false vacuum background far from the bubble. The bubble nucleation rate is eventually well approximated at low $T$ by $\Gamma\equiv \Gamma_4$, where [cf.~Eq.~\eqref{eq:bubble}]
\begin{equation}
    \Gamma_4 \simeq \frac{1}{R_c^4}\left(\frac{S_4}{2\pi}\right)^2\exp(-S_4) \, ,
\end{equation}
with $R_c\sim 1/T$ being the bubble radius in the low $T$ limit.  

At finite temperature, the field becomes periodic in the time coordinate (or in $1/T$). The configuration minimizing the action in this case is $O(3)$-symmetric. Moreover, at sufficiently high temperatures, the minimum action configuration becomes constant in the time direction and $S\equiv S_3(T)/T$, where 
\begin{equation}
S_3(T) = 4\pi \int_0^\infty {\rm d} r \: r^2 \left[\frac{1}{2}\left(\frac{{\rm d}\phi}{\dd r}\right)^2+\Delta V_{\rm eff}(\phi,T)\right] \, ,
\label{eq_S3}
\end{equation}
where $\Delta V_{\rm eff}(\phi,T) \equiv V_{\rm eff}(\phi,T) - V_{\rm eff}(\phi_{\rm false},T)$. Eq.~\eqref{eq_S3} represents bubble formation through classical field excitation over the barrier, with the corresponding equation of motion given by 
\begin{equation}
\frac{{\rm d}^2 \phi}{{\rm d} r^2} + \frac{2}{r} \frac{\dd\phi}{\dd r} - \frac{\partial V_{\rm eff}}{\partial \phi}
= 0 \, ,
\label{eq:saddle}
\end{equation}
with the same boundary conditions as in Eq.~\eqref{eq:boundary}. The solution to Eq.~\eqref{eq:saddle} extremizes the action~\eqref{eq_S3} that gives the exponential suppression of the false vacuum decay rate~\cite{Coleman:1977th}. From Eq.~\eqref{eq:bubble}, the nucleation rate $\Gamma\equiv \Gamma_3$ can be calculated as 
\begin{equation}
\Gamma_3 \simeq T^4\left(\frac{S_3(T)}{2\pi T}\right)^{3/2} \exp\left[-\frac{S_3(T)}{T}\right] \, .
\label{eq_Gamma1}
\end{equation}
In practice, the exact solution with a non-trivial periodic bounce in the time coordinate, which corresponds to quantum tunneling at finite temperature, is difficult to evaluate. Following Ref.~\cite{Espinosa:2018hue}, we have taken the minimum of the two actions $S_3$ and $S_4$ in our numerical calculation of the bubble nucleation rate, {\it i.e.}, $\Gamma\approx {\rm max}(\Gamma_3,\Gamma_4)$. For a discussion of related theoretical uncertainties, see Ref.~\cite{Croon:2020cgk}.

The nucleation temperature is defined as the inverse time of creation of one bubble per Hubble radius, {\it i.e.},
\begin{equation}
    \left.\frac{\Gamma(T)}{H(T)^4}\right|_{T=T_n} =1 \, ,   \label{eq:Tn} 
\end{equation}
where the Hubble expansion rate at temperature $T$ is $H(T)\simeq 1.66\sqrt{g_*} T^2/M_{\rm Pl}$, with $g_*\simeq 110$ being the relativistic degrees of freedom at high temperatures,\footnote{In our numerical analysis, we take into account the temperature-dependence of $g_*$~\cite{Saikawa:2018rcs}.} and $M_{\rm Pl}\simeq 2.4\times 10^{18}$ GeV being the reduced Planck mass. 

The statistical analysis of the subsequent evolution of bubbles in the early Universe is crucial for SFOPT~\cite{Ellis:2018mja}.  The probability
for a given point to remain in the false vacuum is given by $P(T) = \exp[-I(T)]$~\cite{Guth:1979bh, Guth:1981uk, Guth:1982pn,Enqvist:1991xw}, where $I(T)$ is the expected volume of true vacuum bubbles per comoving volume:
\begin{equation}
    I(T) = \frac{4\pi}{3}\int^{T_c}_{T}\frac{\dd T^\prime}{T^{\prime 4}}\frac{\Gamma(T^\prime)}{H(T^\prime)}\left(\int^{T^\prime}_{T}\frac{\dd\tilde{T}}{H(\tilde{T})}\right)^3.
\end{equation}
The change in the physical volume of the false vacuum, ${\cal V}_{\rm false}=a^3({\rm t})P(T)$ (with $a({\rm t})$ being the scale factor and ${\rm t}$ being the time), normalized to the Hubble rate, is given by~\cite{Turner:1992tz} 
\begin{equation}
    \frac{1}{{\cal V}_{\rm false}}\frac{\dd{\cal V}_{\rm false}}{\dd {\rm t}} = H(T)\left(3+T\frac{\dd I(T)}{\dd T}\right) \, .
\end{equation}
We define the percolation temperature $T_p$ as satisfying the condition $I(T_p)=0.34$~\cite{Ellis:2018mja}, while ensuring that the volume of the  false vacuum is decreasing,~ {\it i.e.},~$\dd\log{\cal V}_{\rm false}/\dd {\rm t}<0$, so that percolation is possible despite the exponential expansion of the false vacuum. Numerically, we find that the percolation temperature is only slightly smaller than the nucleation temperature, which in turn is smaller than the critical temperature. Moreover, as expected, both $T_c$ and $T_n\simeq T_p$ grow monotonically, as either of the model parameters, $M_{Z'}$ or $g'$, increases. They are depicted in Figure 2.

\begin{figure}
    \centering
    \includegraphics[width=0.5\linewidth]{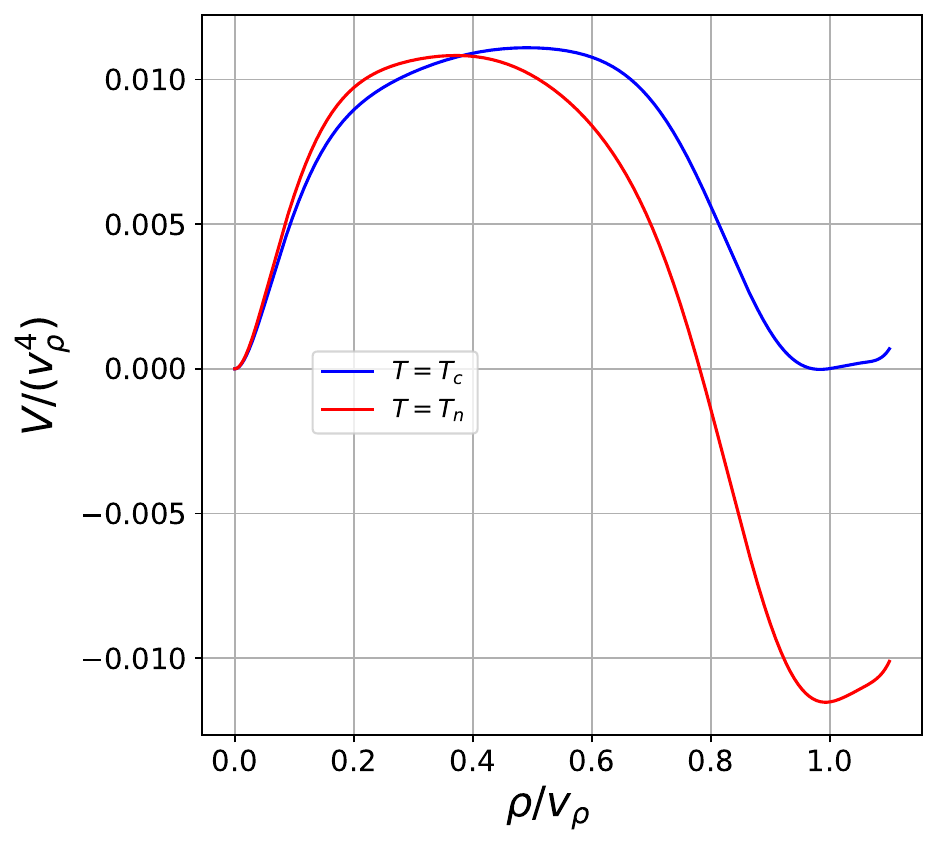}
\caption{ The critical temperature and the nucleation temperature for a typical set of parameters} 
    \label{fig:2}
    \end{figure}

The strength of the phase transition is characterized by two quantities $\alpha$ and $\beta$ defined as follows: $\alpha=\epsilon_*/\rho_{\rm rad}$ is the ratio of the vacuum energy density $\epsilon_*$ released in the transition to the radiation energy density $\rho_{\rm rad}=\pi^2 g_* T_*^4/30$, both evaluated at $T=T_*$ (where $T_*$ is either $T_n$ or $T_p$).\footnote{Calculating $\alpha$ at $T_p$ is better, as it accounts for the entropy dilution  between $T_n$ and $T_p$, although this distinction becomes irrelevant for the GW signal when $\alpha\gg 1$. } The vacuum energy density is nothing but the free energy difference between the true and false vacua~\cite{Espinosa:2008kw}, thus yielding   
\begin{equation}
    \alpha =  \frac{1}{\rho_{\rm rad}}\left.\left(-1+T\frac{\dd}{\dd T}\right)\Delta V_{\min}\right|_{T=T_*} \, ,
    \label{eq:alpha}
\end{equation}
where $\Delta V_{\rm min}$ is the temperature-dependent minimum of the effective potential $\Delta V_{\rm eff}$ defined below Eq.~\eqref{eq_S3}. 

The second important parameter is $\beta/H_*$, where $\beta$ is the (approximate) inverse timescale of the phase transition and $H_*$ is the Hubble rate at $T_*$:
 \begin{equation}
     \frac{\beta}{H_*} = \left. -\frac{T}{\Gamma} \frac{\dd\Gamma}{\dd T}\right|_{T=T_*} \, .
     \label{eq:beta}
 \end{equation}
 For strong transitions, $\beta$ is related to the average bubble radius $R_*$: $\beta=(8\pi)^{1/3}/R_*$~\cite{Ellis:2020nnr},\footnote{When identifying $\beta/H_*$ with the average bubble radius, it is numerically found that taking $T_*=T_n$ gives a more accurate result~\cite{Baldes:2023rqv}.} where $R_*$ defines the characteristic length scale of transition and is given by~\cite{Turner:1992tz, Enqvist:1991xw}
 \begin{equation}
     R_* = \left[T_*\int_{T_*}^{T_c}\frac{\dd T}{T^2}\frac{\Gamma(T)}{H(T)}\exp\{-I(T)\}\right]^{-1/3} \, .
 \end{equation}

The amplitude of the GW signal as a function of the frequency $f$ is usually defined as 
\begin{equation}
h^2\Omega_{\rm GW}(f)\equiv \frac{h^2}{\rho_c}\frac{\dd\rho_{\rm GW}}{\dd\log f} \, ,
\end{equation}
where $h\sim 0.7$ is the dimensionless Hubble parameter (defined in terms of today's value of $H$, $H_0=100 h~{\rm km/s/Mpc}$), $\rho_{\rm GW}$ is the energy density released in the form of GWs, and $\rho_c=3H_0^2M_{\rm Pl}^2\simeq 1.05\times 10^{-5}h^2~{\rm GeV}/{\rm cm}^3$ is the critical density of the Universe. The reason for multiplying $\Omega_{\rm GW}$ by $h^2$ is to make sure that the GW amplitude is not affected by the experimental uncertainty~\cite{DiValentino:2021izs} in the Hubble parameter $H_0$. 

There are three different mechanisms for producing GWs in an SFOPT from the expanding and colliding scalar-field bubbles,  as well as from their interaction with the thermal plasma. These are: (i) collisions of expanding bubble walls~\cite{Kosowsky:1991ua,Kosowsky:1992rz,Kosowsky:1992vn,Kamionkowski:1993fg,Caprini:2007xq,Huber:2008hg,Weir:2016tov, Jinno:2017fby, Jinno:2019bxw,Lewicki:2020jiv,Megevand:2021juo}, compressional modes (or sound waves) in the bulk plasma~\cite{Hindmarsh:2013xza,Giblin:2013kea,Giblin:2014qia,Hindmarsh:2015qta, Hindmarsh:2016lnk,Hindmarsh:2017gnf,Hindmarsh:2019phv}, and (iii) vortical motion (or magnetohydrodynamic turbulence) in the bulk plasma~\cite{Caprini:2006jb,Kahniashvili:2008pf,Kahniashvili:2008pe,Kahniashvili:2009mf,Caprini:2009yp,Kisslinger:2015hua, RoperPol:2019wvy}. The total GW signal can be approximated as a linear superposition of the signals generated from these three individual sources, denoted respectively by $\Omega_b$ (bubble wall), $\Omega_s$ (sound wave), and $\Omega_t$ (turbulence):
\begin{equation}
    h^2\Omega_{\rm GW}(f) \simeq h^2\Omega_b(f) + h^2\Omega_s (f) + h^2\Omega_t(f) \, .
    \label{eq:tot_gw}
\end{equation}
The three contributions can be parameterized in a model-independent way in terms of a set of characteristic SFOPT parameters, namely,  $\alpha$ [cf.~Eq.~\eqref{eq:alpha}], $\beta/H_*$ [cf.~Eq.~\eqref{eq:beta}], $T_*$,\footnote{We will use $T_*=T_n$, the nucleation temperature defined in Eq.~\eqref{eq:Tn}. For subtleties, see Ref.~\cite{Athron:2022mmm}.} bubble-wall velocity $v_w$, and the three efficiency factors $\kappa_b$, $\kappa_s$,  $\kappa_t$ that characterize the fractions of the released vacuum energy that are converted into the energy of scalar field gradients, sound waves and turbulence,  respectively. The bubble-wall velocity in the plasma rest-frame is given by~\cite{Lewicki:2021pgr} 
\begin{equation}
    v_w = \left\{\begin{array}{ll}\sqrt{\frac{\Delta V_{\rm min}}{\alpha\rho_{\rm rad}}} & {\rm for}~\frac{\Delta V_{\rm min}}{\alpha\rho_{\rm rad}}<v_J \\
    1 & {\rm for}~\frac{\Delta V_{\rm min}}{\alpha\rho_{\rm rad}}\geq v_J \end{array}\right. \, ,
\end{equation}
where $v_J=(1+\sqrt{3\alpha^2+2\alpha})/\sqrt 3(1+\alpha)$ is the Jouguet velocity~\cite{Kamionkowski:1993fg, Steinhardt:1981ct,Espinosa:2010hh}. As for the efficiency factors, it is customary to express $\kappa_s$ and $\kappa_t$ in terms of another efficiency factor $\kappa_{\rm kin}$ that characterizes the energy fraction converted into bulk kinetic energy and an additional parameter $\varepsilon$, {\it i.e.}~\cite{Ellis:2019oqb} 
\begin{equation}
    \kappa_s = \kappa_{\rm kin} \, , \qquad \kappa_t = \varepsilon \kappa_{\rm kin} \, .
\end{equation}
While the precise numerical value of $\varepsilon$ is still under debate, following Refs.~\cite{Caprini:2015zlo,Hindmarsh:2015qta}, we will use $\varepsilon=1$. 
The efficiency factor $\kappa_b$ is taken from Ref.~\cite{Kamionkowski:1993fg} and $\kappa_{\rm kin}$ is taken from Ref.~\cite{Espinosa:2010hh}, both of which were calculated in the so-called Jouguet detonation limit:
\begin{equation}
    \begin{aligned}
        \kappa_b &= \frac{1}{(1 + 0.715 \alpha)}\left(0.715 \alpha + \frac{4}{27} \sqrt{\frac{3\alpha}{2}}\right) \, ,  \\
        \kappa_{\rm kin} &= \frac{\sqrt{\alpha}}{(0.135 + \sqrt{0.98 + \alpha})} \, .
    \end{aligned}
\end{equation}
Each of the three contributions in Eq.~\eqref{eq:tot_gw} is related to the SFOPT parameters discussed above, as follows~\cite{Schmitz:2020syl}:
\begin{equation}
    h^2 \Omega_i(f) = h^2\Omega^{\rm peak}_i\left(\alpha,\frac{\beta}{H_*},T_*,v_w,\kappa_i\right)\mathcal{S}_i(f,f_i) \, , 
    \label{eq:Omega}
\end{equation}
where $i\in\{b,s,t\}$, the peak amplitudes are given as~\cite{Caprini:2015zlo} 
\begin{equation}
\begin{aligned}
    h^2\Omega^{\rm peak}_b &\simeq 1.67\times 10^{-5} \left(\frac{v_w}{\beta/H_*}\right)^2\left(\frac{100}{g_*(T_*)}\right)^{1/3}\left(\frac{\kappa_b \alpha}{1+\alpha}\right)^2\left(\frac{0.11v_w}{0.42 + v^2_w}\right), \\
    h^2\Omega^{\rm peak}_s &\simeq 2.65\times 10^{-6} \left(\frac{v_w}{\beta/H_*}\right)\left(\frac{100}{g_*(T_*)}\right)^{1/3}\left(\frac{\kappa_s \alpha}{1+\alpha}\right)^2,  \\
    h^2\Omega^{\rm peak}_t &\simeq 3.35\times 10^{-4} \left(\frac{v_w}{\beta/H_*}\right)\left(\frac{100}{g_*(T_*)}\right)^{1/3}\left(\frac{\kappa_t \alpha}{1+\alpha}\right)^{3/2},
    \label{eq:Opeak}
\end{aligned}
\end{equation}
and the spectral shape functions are given as~\cite{Caprini:2015zlo} 
\begin{equation}
\begin{aligned}
    \mathcal{S}_b(f,f_b) &= \left(\frac{f}{f_b}\right)^{2.8}\left[\frac{3.8}{1+2.8(f/f_b)^{3.8}}\right],  \\
    \mathcal{S}_s(f,f_s) &= \left(\frac{f}{f_s}\right)^3\left[\frac{7}{4+3(f/f_s)^2}\right]^{7/2},  \\
    \mathcal{S}_t(f,f_t,h_*) &= \left(\frac{f}{f_t}\right)^3\left[\frac{1}{1+(f/f_t)}\right]^{11/3}\left(\frac{1}{1+8\pi f/h_*}\right)\, .
\end{aligned}
\end{equation}
Note that ${\cal S}_b$ and ${\cal S}_s$ are normalized to unity at their respective peak frequencies $f_b$ and $f_s$, whereas ${\cal S}_t$ at $f_t$ depends on the Hubble frequency 
\begin{equation}
    h_* = \frac{a_*}{a_0}H_*=1.6\times 10^{-2}~\textrm{mHz} \left(\frac{g_*(T_*)}{100}\right)^{1/6}\left(\frac{T_*}{100~\gev}\right).
\end{equation}
\begin{figure}
    \centering
    \includegraphics[width=0.7\linewidth]{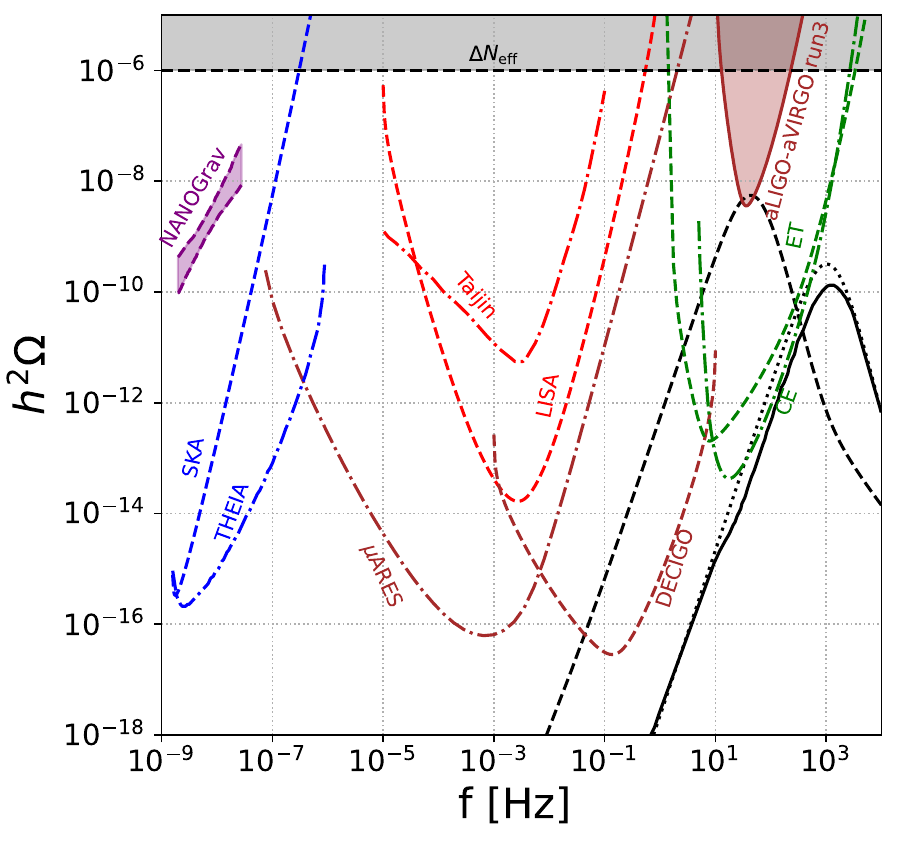}
     
\caption{The gravitational wave spectrum for  our benchmark points.  Benchmark point 1 is the solid line, benchmark point 2 is the dashed line and benchmark point 3 is the dotted line.  Note that since the electroweak transition temperature is much higher than in most models, the peak frequency is also substantially higher. CE and ET are the expected sensitivity of the Cosmic Explorer and Einstein Telescope, respectively.} 
    \label{fig:3}
    \end{figure}

Finally, the peak frequencies are given as
\begin{equation}
\begin{aligned}
    f_b &= 1.6\times10^{-2}~\textrm{mHz} \left(\frac{g_*(T_*)}{100}\right)^{1/6}\left(\frac{T_*}{100~\gev}\right)\left(\frac{\beta/H_*}{v_w}\right)\left(\frac{0.62 v_w}{1.8 - 0.1 v_w + v^2_w}\right),  \\
    f_s &= 1.9\times10^{-2}~\textrm{mHz} \left(\frac{g_*(T_*)}{100}\right)^{1/6}\left(\frac{T_*}{100~\gev}\right)\left(\frac{\beta/H_*}{v_w}\right),  \\
    f_t &= 2.7\times10^{-2}~\textrm{mHz} \left(\frac{g_*(T_*)}{100}\right)^{1/6}\left(\frac{T_*}{100~\gev}\right)\left(\frac{\beta/H_*}{v_w}\right). \label{eq:fpeak}
\end{aligned}
\end{equation}

The results are shown for our benchmark points in Figure 3.   Many gravitational wave signals from electroweak transitions are in the LISA/DECIGO range, but here we see that the frequency of the gravitational wave signal is higher, primarily due to the larger transition temperature.    The signal would not have been seen by LIGO but might be in range of future gravitational wave detectors in that frequency range.

\section{Transport Equation and Asymmetry}

Since sphalerons conserve $B-L$, a popular mechanism of baryogenesis is to generate a lepton asymmetry which sphalerons convert to a baryon asymmetry.   One scenario for lepton asymmetry is a right-handed Majorana neutrino mass\cite{Davidson:2002qv}.   This has advantages and disadvantages - it does not require CP violation in the quark sector which may alleviate constraints from EDM experiments, but it also may not have any collider signatures since such a lepton asymmetry from a right-handed neutrino mass will typically arise at a very high scale.   Although that could also work here, the fact that we have a strongly first order phase transition makes it tempting to study whether the baryon asymmetry can be directly generated in this model.   To this end, a complex phase must occur in the Higgs sector.

In a follow-up article to the Gu model described above\footnote{In Gu's work, a lepton asymmetry is generated through a heavy Majorana neutrino, which is then converted to a baryon asymmetry.}, Gu and Lindner\cite{Gu:2010zv} added a complex singlet field $\xi$.   The purpose of that work was to include a Peccei-Quinn symmetry.   For our purposes, it is sufficient to include the phase of $\xi$.   None of our previous results will be substantially changed by addition of this field.     {This is because the fact that $\xi$ does not get a vev implies that its only effect is on undetermined parameters and masses}

The baryon asymmetry for our model is produced at the bubble walls specifically when $v_R = v_L \neq 0$ or when $v_L$ gets the vev. We calculate this asymmetry using the method presented in \cite{Cline:2020jre}.

\begin{figure}[!h]
    \centering
    \begin{tabular}{lr}
      \begin{tikzpicture}[/tikzfeynman/small]
      \begin{feynman}
      \vertex (v1);
      \vertex[above left = 1. cm of v1](i){$H_L$};
      \vertex[below left = 1. cm of v1](j){$f$};
      \vertex[right= 2.cm of v1](v2);
      \vertex[above right= 1.cm of v2](k){$H_R$};
      \vertex[below right = 1.cm of v2](l){$\bar{f}$};
      \diagram*[small]{(j)--[fermion](v1)--[scalar](i),(v1)--[anti majorana,edge label'=$\bar{F} \quad F$](v2),(k)--[scalar](v2)--[anti fermion](l)};
      \end{feynman}   
      \end{tikzpicture} &
      \end{tabular}
    \caption{Feynman diagrams for generating the baryon asymmetry.}
    \label{fig:Qt}
\end{figure}
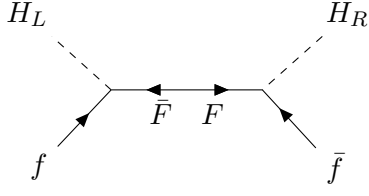

For this method, transport equations are used which describe how particle asymmetries are transferred across bubble walls. The Boltzmann equations derived use the zeroth and first order moment expansions of the Louiville operator to produce two first-order differential equations per particle species relating each species' \emph{CP-odd} chemical potential, $\mu_i$, and their velocity perturbations, $u_i$. Any coupling between particle species arises from the collision term $\delta C_i$. The collision terms include contributions from trilinear and quartic terms. The relevant Feynman diagrams are shown in Figure \ref{fig:Qt}.

The main source term for the $CP$ violation comes form the mass term for each fermion which is given as:
\begin{align}
    m_f = y^2_f \frac{v_L v_R}{M_f}
\end{align}
 Furthermore, the complex part comes from $M_f$ from the $\xi$ field mentioned earlier. The relevant part comes from 
   {
 }
 \begin{align}
 \mathcal{L}_F &\supset (M_F + iy_\xi \xi)\overline{F}F  \\
     M_f & = M_F + i y_{\xi f}  \xi  = \widetilde{M}_fe^{i\theta_f} \quad \theta_f = \tan^{-1}\left(\frac{y_{\xi f} \xi }{M_f}\right)\\
     m_f &= y^2 \frac{v_Rv_L}{\widetilde{M_f}}e^{-i\theta_f}\\
     m^2_f &= y^4 \frac{v^2_Rv^2_L}{\widetilde{M_f}^2}e^{-2i\theta_f}
 \end{align}
As noted, the $\xi$ is an additional field whose only purpose is to provide us with the relevant complex phase.  We have the freedom to choose the coupling of this scalar with the $H_L$ such that the vev of $\langle \xi \rangle \neq 0$ around the transition temperature $T_L$. Now to analyze the phase transition we closely follow \cite{Espinosa:2011eu} and choose the kink profile for the scalars $H_L,\xi$ given as:
\begin{align}
    \xi(z) &= \frac{\langle \xi \rangle}{2}\left[1 + \tanh{\left(\frac{z}{L_w}\right)}\right]; \\
    H_L(z) &= \frac{\langle H_L \rangle}{2}\left[ 1 + \tanh{\left(\frac{z}{L_w}\right)}\right] \\
    L^2_w &\simeq 2.7 \times \frac{1}{\kappa}\frac{v^2_L + v^2_\xi}{v_Lv_\xi}\left(1+ \frac{\kappa v^2_\xi}{4\lambda_H v^2_L}\right).  
\end{align}
where $L_w$ is the length of the bubble profile and $\kappa$ is the effective coupling between the scalar fields ($\mathcal{L} \supset \frac{\kappa}{4}  \xi^2 H^2_L)$.

The source term $S_f$ is given as\cite{Cline:1995dg} \footnote{The expression of $Q^{8o}_l$, $Q^{9o}_l$ and $D_0$ are mentioned in \cite{Cline:1995dg}.}:
\begin{align}
    S_{fl} &= -v_w\gamma_w h\left[(m_f^2\theta^\prime)^\prime Q^{8o}_l - (m_f^2)^\prime m_f^2 \theta^\prime Q^{9o}_l\right], \\
\end{align}

The transport equations use the WKB approximation and treat the particle interactions as a semi-classical force. For this to be a valid approximation, the thickness of the bubble wall, $L$, should be greater than $T^{-1}$. 

\be
A_iw_i' + \left(m_i^2\right)'B_iw_i = S_i + \delta C_i
\ee
where,
\begin{align}
    A_i &= \begin{pmatrix}
        -D_1 & 1 \\ -D2 & R
    \end{pmatrix}, \quad B_i = \begin{pmatrix}
        v_w \gamma_w Q_1 & 0 \\ v_w\gamma_w Q_2 & \bar{R}
    \end{pmatrix}
\end{align}
all the integrals $D_i,R$ and $\bar{R}$ is given in \cite{Cline:1995dg}.

The collision terms include Yukawa interactions, Yukawa interactions with an external gluon line, strong sphaleron rate, and $W_L$/$W_R$ interactions. The temperature of the phase transition is much lower than the mass of the heavy fermions in our model, so the Yukawa interactions involving these and the SM fermions are kinematically forbidden. The relevant terms are shown below.
\begin{table}[!h]
    \centering
    \begin{tabular}{|c|c|c|}
      \hline \hline & BP2  & BP3 \\
     \hline  $\langle \xi \rangle$ & 8 TeV &  18 TeV\\
       $\kappa$ & 0.005 & 0.0001\\
       $\eta_B/\eta_{B_{obs}}$ & 1.5 & 1.8\\
       \hline
    \end{tabular}
    \caption{The baryon asymmetry for BP2 and BP3 (BP1 is not phenomenologically acceptable so we do not consider it here).    Each value of $\eta_B/\eta_{B_{obs}}$ should be multiplied by $\sin\arg(\xi)$.}
    \label{tab:my_label}
\end{table}
\begin{align}
    \delta C_1^t &= \Gamma_{m_t}\left(\mu_t - \mu_{\bar{t}}\right) + \Gamma_{W_L}\left(\mu_t-\mu_b\right) + \tilde{\Gamma}_{SS}\left[\mu_i\right] + \tilde{\Gamma}_{W_L}\left[\mu_i\right] \\
    \delta C_1^{\bar{t}} &= \Gamma_{m_t}\left(\mu_{\bar{t}} - \mu_{t}\right) + \Gamma_{W_R}\left(\mu_{\bar{t}}-\mu_{\bar{b}}\right) - \tilde{\Gamma}_{SS}\left[\mu_i\right] + \tilde{\Gamma}_{W_R}\left[\mu_i\right] \\
    \delta C_1^b &= \Gamma_{m_b}\left(\mu_b - \mu_{\bar{b}}\right) + \Gamma_{W_L}\left(\mu_b-\mu_t\right) + \tilde{\Gamma}_{SS}\left[\mu_i\right] + \tilde{\Gamma}_{W_L}\left[\mu_i\right] \\
    \delta C_1^{\bar{b}} &= \Gamma_{m_b}\left(\mu_{\bar{b}} - \mu_{b}\right) + \Gamma_{W_R}\left(\mu_{\bar{b}} - \mu_{\bar{t}}\right) - \tilde{\Gamma}_{SS}\left[\mu_i\right] + \tilde{\Gamma}_{W_R}\left[\mu_i\right] \\
    \delta C_1^{h_L} &= \Gamma_{h_L}\mu_{h_L} \\
    \delta C_1^{h_R} &= \Gamma_{h_R} \mu_{h_R} 
\end{align}

To solve these coupled differential equations, we used relaxation methods that approximate the differential equations as stiff equations. After the chemical potential across the bubble wall is calculated using this method, the baryon asymmetry is calculated using the method of Ref. \cite{Prokopec:2013ax}.

\begin{align}
\eta_B &= \frac{405\Gamma_{sph}}{4\pi^2v_w\gamma_wg_*T}\int dz\mu_{B_L} f_{sph}e^{-45\Gamma_{sph}|z|/4v_w\gamma_w}  \\
\mu_{B_L} &= \frac{1}{2}(1 + 4D^t_0)\mu_t + \frac{1}{2}(1 + 4D^b_0)\mu_b + 2D^{\bar{t}}_0\mu_{\bar{t}}
\end{align}

$h\sigma(\bar{D_L}D_R) + {\rm h.c} = h\sigma(\bar{D_L}D_R)-h\sigma(\bar{D_R}D_L)  =
h\sigma(\bar{D_L}D_R - \bar{D_R}D_L)$

\be
\tilde{\Gamma}_{W_{R,L}}\left[\mu_i\right] = \Gamma_{\tilde{W}_{R,L}}\left(\left(1+9D_0^{\bar{t},t}\right)\mu_{\bar{t},t} + \left(1+9D_0^{\bar{b},b}\right)\mu_{\bar{b},b}\right)
\ee

\be
\tilde{\Gamma}_{SS}\left[\mu_i\right] = \Gamma_{SS}\left(\left(1+9D_0^t\right)\mu_t + \left(1+9D_0^b\right)\mu_b - \left(1-9D_0^{\bar{t}}\right)\mu_{\bar{t}}\right)
\ee

Our results for BP2 and BP3 are in Table 3.   The vev of $\xi$ is large, but since it is a gauge singlet it will not affect the results of the previous section.    For each point, the baryon asymmetry is in the acceptable range, depending on the phase of $\xi$.

For BP1,  $v_L=v_R\neq 0$  earlier in time. Now, in this case the asymmetry will be created at that epoch $T_{LR}$ (the $\xi$ gets a vev at the same time creating the $CP$ violation). At the same epoch the sphaleron of both left and right will go out-of-equilibrium and hence the asymmetry created at that epoch will be frozen in.  For this scenario one might need to consider right-handed sphalerons, but that isn't necessary since BP1 is already excluded by the $W_R$ mass.

\section{Conclusions}

One of the most exciting discoveries of the past decade was the discovery of gravitational waves.   Future experiments are planned that will provide an extraordinary increase in sensitivity.   It had long been believed that the only experimental information about the electroweak phase transition was the single number giving the baryon asymmetry.  Now, however, the gravitational wave spectrum might give much more information.  A crucial input to understanding the spectrum is the temperature of the electroweak transition.   We have considered a left-right model with two Higgs doublets and a singlet (although more complicated Higgs structures are possible) and shown that the transition temperature can be much higher than the current electroweak scale.   This leads to a spectrum that peaks at higher frequencies than in previous studies, possibly in the range of the Cosmic Explorer or the Einstein Telescope.
We are not tied to a specific model and feel that this result might be generic.  Certainly if a right-handed W is discovered at the LHC, a much more detailed analysis will be warranted.  We also discuss electroweak baryogenesis in the model and show that the phenomenologically acceptable benchmark points can easily give the necessary baryon asymmetry if a complex singlet is added to provide a CP violating phase.

\acknowledgments

The work of MK and MS is supported by the National Science Foundation under Grants PHY-2112460 and (for MK) PHY-2411549.   MK thanks the Pittsburgh Particle Physics, Astrophysics and Cosmology Center (PITT-PACC) and the University of Pittsburgh for their hospitality.   The work of AD was supported in part by the U.S. Department of Energy under grant No. DE-SC0007914 and in part by PITT-PACC.

\newpage
\bibliographystyle{apsrev}


\begin{thebibliography}{99}
\bibitem{Planck:2018vyg}
N.~Aghanim \textit{et al.} [Planck],
Astron. Astrophys. \textbf{641}, A6 (2020)
[erratum: Astron. Astrophys. \textbf{652}, C4 (2021)]
[arXiv:1807.06209 [astro-ph.CO]].
\bibitem{kuzmin}
V. A. Kuzmin, V. A. Rubakov and M. E. Shaposhnikov,
Phys. Lett. \textbf{155B}, 36 (1985)
\bibitem{Fukugita:1986hr}
M.~Fukugita and T.~Yanagida,
Phys. Lett. B \textbf{174} (1986), 45-47
\bibitem{Cohen:1993nk}
A.~G.~Cohen, D.~B.~Kaplan and A.~E.~Nelson,
Ann. Rev. Nucl. Part. Sci. \textbf{43} (1993), 27-70
[arXiv:hep-ph/9302210 [hep-ph]].
\bibitem{Rubakov:1996vz}
V.~A.~Rubakov and M.~E.~Shaposhnikov,
Usp. Fiz. Nauk \textbf{166} (1996), 493-537
[arXiv:hep-ph/9603208 [hep-ph]].
\bibitem{Trodden:1998ym}
M.~Trodden,
Rev. Mod. Phys. \textbf{71} (1999), 1463-1500
[arXiv:hep-ph/9803479 [hep-ph]].
\bibitem{Morrissey:2012db}
D.~E.~Morrissey and M.~J.~Ramsey-Musolf,
New J. Phys. \textbf{14} (2012), 125003
[arXiv:1206.2942 [hep-ph]].
\bibitem{Sakharov}
A.D. Sakharov, Pisma Zh. Eksp. Teor Fiz. \textbf{5}, 32 (1967)
\bibitem{Rummukainen:1998as}
K.~Rummukainen, M.~Tsypin, K.~Kajantie, M.~Laine and M.~E.~Shaposhnikov,
Nucl. Phys. B \textbf{532}, 283-314 (1998)
doi:10.1016/S0550-3213(98)00494-5
[arXiv:hep-lat/9805013 [hep-lat]].
\bibitem{Branco:2011iw}
G.~C.~Branco, P.~M.~Ferreira, L.~Lavoura, M.~N.~Rebelo, M.~Sher and J.~P.~Silva,
Phys. Rept. \textbf{516}, 1-102 (2012)
[arXiv:1106.0034 [hep-ph]].
\bibitem{Bochkarev:1990fx}
A.~I.~Bochkarev, S.~V.~Kuzmin and M.~E.~Shaposhnikov,
Phys. Lett. B \textbf{244} (1990), 275-278
\bibitem{Turok:1990zg}
N.~Turok and J.~Zadrozny,
Nucl. Phys. B \textbf{358} (1991), 471-493
\bibitem{Turok:1991uc}
N.~Turok and J.~Zadrozny,
Nucl. Phys. B \textbf{369} (1992), 729-742
\bibitem{Funakubo:1993jg}
K.~Funakubo, A.~Kakuto and K.~Takenaga,
Prog. Theor. Phys. \textbf{91} (1994), 341-352
[arXiv:hep-ph/9310267 [hep-ph]].
\bibitem{Davies:1994id}
A.~T.~Davies, C.~D.~froggatt, G.~Jenkins and R.~G.~Moorhouse,
Phys. Lett. B \textbf{336} (1994), 464-470
\bibitem{Funakubo:1995kw}
K.~Funakubo, A.~Kakuto, S.~Otsuki, K.~Takenaga and F.~Toyoda,
Prog. Theor. Phys. \textbf{94} (1995), 845-860
[arXiv:hep-ph/9507452 [hep-ph]].
\bibitem{Funakubo:1996iw}
K.~Funakubo, A.~Kakuto, S.~Otsuki and F.~Toyoda,
Prog. Theor. Phys. \textbf{96} (1996), 771-780
[arXiv:hep-ph/9606282 [hep-ph]].
\bibitem{Cline:1995dg}
J.~M.~Cline, K.~Kainulainen and A.~P.~Vischer,
Phys. Rev. D \textbf{54} (1996), 2451-2472
[arXiv:hep-ph/9506284 [hep-ph]].
\bibitem{Fuyuto:2015jha}
K.~Fuyuto and E.~Senaha,
Phys. Lett. B \textbf{747} (2015), 152-157
[arXiv:1504.04291 [hep-ph]].
\bibitem{Chiang:2016vgf}
C.~W.~Chiang, K.~Fuyuto and E.~Senaha,
Phys. Lett. B \textbf{762} (2016), 315-320
[arXiv:1607.07316 [hep-ph]].
\bibitem{Dorsch:2013wja}
G.~C.~Dorsch, S.~J.~Huber and J.~M.~No,
JHEP \textbf{10} (2013), 029
[arXiv:1305.6610 [hep-ph]].
\bibitem{Dorsch:2014qja}
G.~C.~Dorsch, S.~J.~Huber, K.~Mimasu and J.~M.~No,
Phys. Rev. Lett. \textbf{113} (2014) no.21, 211802
[arXiv:1405.5537 [hep-ph]].
\bibitem{Andersen:2017ika}
J.~O.~Andersen, T.~Gorda, A.~Helset, L.~Niemi, T.~V.~I.~Tenkanen, A.~Tranberg, A.~Vuorinen and D.~J.~Weir,
Phys. Rev. Lett. \textbf{121}, no.19, 191802 (2018)
[arXiv:1711.09849 [hep-ph]].
\bibitem{Bernon:2017jgv}
J.~Bernon, L.~Bian and Y.~Jiang,
JHEP \textbf{05}, 151 (2018)
[arXiv:1712.08430 [hep-ph]].
\bibitem{Kainulainen:2019kyp}
K.~Kainulainen, V.~Keus, L.~Niemi, K.~Rummukainen, T.~V.~I.~Tenkanen and V.~Vaskonen,
JHEP \textbf{06}, 075 (2019)
[arXiv:1904.01329 [hep-ph]].
\bibitem{Bittar:2025lcr}
P.~Bittar, S.~Roy and C.~E.~M.~Wagner,
[arXiv:2504.02024 [hep-ph]].

\bibitem{Basler:2016obg}
P.~Basler, M.~Krause, M.~Muhlleitner, J.~Wittbrodt and A.~Wlotzka,
JHEP \textbf{02}, 121 (2017)
[arXiv:1612.04086 [hep-ph]].

\bibitem{Cline:1996mga}
J.~M.~Cline and P.~A.~Lemieux,
Phys. Rev. D \textbf{55} (1997), 3873-3881
[arXiv:hep-ph/9609240 [hep-ph]].
\bibitem{Fromme:2006cm}
L.~Fromme, S.~J.~Huber and M.~Seniuch,
JHEP \textbf{11} (2006), 038
[arXiv:hep-ph/0605242 [hep-ph]].
\bibitem{Cline:2011mm}
J.~M.~Cline, K.~Kainulainen and M.~Trott,
JHEP \textbf{11} (2011), 089
[arXiv:1107.3559 [hep-ph]].
\bibitem{Dorsch:2016nrg}
G.~C.~Dorsch, S.~J.~Huber, T.~Konstandin and J.~M.~No,
JCAP \textbf{05} (2017), 052
[arXiv:1611.05874 [hep-ph]].
\bibitem{Haarr:2016qzq}
A.~Haarr, A.~Kvellestad and T.~C.~Petersen,
[arXiv:1611.05757 [hep-ph]].
\bibitem{Basler:2017uxn}
P.~Basler, M.~M\"uhlleitner and J.~Wittbrodt,
JHEP \textbf{03}, 061 (2018)
[arXiv:1711.04097 [hep-ph]].
\bibitem{Wang:2019pet}
X.~Wang, F.~P.~Huang and X.~Zhang,
Phys. Rev. D \textbf{101}, no.1, 015015 (2020)
[arXiv:1909.02978 [hep-ph]].
\bibitem{Enomoto:2021dkl}
K.~Enomoto, S.~Kanemura and Y.~Mura,
JHEP \textbf{01}, 104 (2022)
doi:10.1007/JHEP01(2022)104
[arXiv:2111.13079 [hep-ph]].


\bibitem{Basler:2021kgq}
P.~Basler, L.~Biermann, M.~M\"uhlleitner and J.~M\"uller,
Eur. Phys. J. C \textbf{83} (2023) no.1, 57
[arXiv:2108.03580 [hep-ph]].



\bibitem{Apreda:2001us}
R.~Apreda, M.~Maggiore, A.~Nicolis and A.~Riotto,
Nucl. Phys. B \textbf{631}, 342-368 (2002)
[arXiv:gr-qc/0107033 [gr-qc]].
\bibitem{Grojean:2006bp}
C.~Grojean and G.~Servant,
Phys. Rev. D \textbf{75}, 043507 (2007)
[arXiv:hep-ph/0607107 [hep-ph]].
\bibitem{Leitao:2012tx}
L.~Leitao, A.~Megevand and A.~D.~Sanchez,
JCAP \textbf{10} (2012), 024
[arXiv:1205.3070 [astro-ph.CO]].
\bibitem{Hindmarsh:2013xza}
M.~Hindmarsh, S.~J.~Huber, K.~Rummukainen and D.~J.~Weir,
Phys. Rev. Lett. \textbf{112} (2014), 041301
[arXiv:1304.2433 [hep-ph]].
\bibitem{Kakizaki:2015wua}
M.~Kakizaki, S.~Kanemura and T.~Matsui,
Phys. Rev. D \textbf{92}, no.11, 115007 (2015)
doi:10.1103/PhysRevD.92.115007
[arXiv:1509.08394 [hep-ph]].
\bibitem{Chala:2018opy}
M.~Chala, M.~Ramos and M.~Spannowsky,
Eur. Phys. J. C \textbf{79}, no.2, 156 (2019)
doi:10.1140/epjc/s10052-019-6655-1
[arXiv:1812.01901 [hep-ph]].
\bibitem{Morais:2019fnm}
A.~P.~Morais and R.~Pasechnik,
JCAP \textbf{04}, 036 (2020)
doi:10.1088/1475-7516/2020/04/036
[arXiv:1910.00717 [hep-ph]].
\bibitem{Caprini:2019egz}
C.~Caprini, M.~Chala, G.~C.~Dorsch, M.~Hindmarsh, S.~J.~Huber, T.~Konstandin, J.~Kozaczuk, G.~Nardini, J.~M.~No and K.~Rummukainen, \textit{et al.}
JCAP \textbf{03}, 024 (2020)
[arXiv:1910.13125 [astro-ph.CO]].
\bibitem{Goncalves:2021egx}
D.~Gon\c{c}alves, A.~Kaladharan and Y.~Wu,
Phys. Rev. D \textbf{105}, no.9, 095041 (2022)
[arXiv:2108.05356 [hep-ph]].


\bibitem{Baldes:2021vyz}
I.~Baldes, S.~Blasi, A.~Mariotti, A.~Sevrin and K.~Turbang,
Phys. Rev. D \textbf{104}, no.11, 115029 (2021)
[arXiv:2106.15602 [hep-ph]].
\bibitem{Azatov:2021irb}
A.~Azatov, M.~Vanvlasselaer and W.~Yin,
JHEP \textbf{10}, 043 (2021)
[arXiv:2106.14913 [hep-ph]].
\bibitem{Benincasa:2022elt}
N.~Benincasa, L.~Delle Rose, K.~Kannike and L.~Marzola,
JCAP \textbf{12}, 025 (2022)
doi:10.1088/1475-7516/2022/12/025
[arXiv:2205.06669 [hep-ph]].
\bibitem{Huang:2022vkf}
P.~Huang and K.~P.~Xie,
JHEP \textbf{09}, 052 (2022)
[arXiv:2206.04691 [hep-ph]].
\bibitem{Dasgupta:2022isg}
A.~Dasgupta, P.~S.~B.~Dev, A.~Ghoshal and A.~Mazumdar,
Phys. Rev. D \textbf{106}, no.7, 075027 (2022)
[arXiv:2206.07032 [hep-ph]].
\bibitem{Cao:2022ocg}
Q.~H.~Cao, K.~Hashino, X.~X.~Li and J.~H.~Yue,
[arXiv:2212.07756 [hep-ph]].
\bibitem{Chatterjee:2022pxf}
A.~Chatterjee, A.~Datta and S.~Roy,
JHEP \textbf{06}, 108 (2022)
doi:10.1007/JHEP06(2022)108
[arXiv:2202.12476 [hep-ph]].

\bibitem{Ghosh:2022fzp}
P.~Ghosh, T.~Ghosh and S.~Roy,
JHEP \textbf{10}, 057 (2023)
doi:10.1007/JHEP10(2023)057
[arXiv:2211.15640 [hep-ph]].

\bibitem{Huang:2016odd}
F.~P.~Huang, Y.~Wan, D.~G.~Wang, Y.~F.~Cai and X.~Zhang,
Phys. Rev. D \textbf{94} (2016) no.4, 041702
[arXiv:1601.01640 [hep-ph]].
\bibitem{Beniwal:2017eik}
A.~Beniwal, M.~Lewicki, J.~D.~Wells, M.~White and A.~G.~Williams,
JHEP \textbf{08} (2017), 108
[arXiv:1702.06124 [hep-ph]].
\bibitem{Demidov:2017lzf}
S.~V.~Demidov, D.~S.~Gorbunov and D.~V.~Kirpichnikov,
Phys. Lett. B \textbf{779} (2018), 191-194
[arXiv:1712.00087 [hep-ph]].
\bibitem{Goncalves:2023svb}
D.~Gon\c{c}alves, A.~Kaladharan and Y.~Wu,
Phys. Rev. D \textbf{108}, no.7, 075010 (2023)
doi:10.1103/PhysRevD.108.075010
[arXiv:2307.03224 [hep-ph]].



\bibitem{Zhou:2020irf}
R.~Zhou and L.~Bian,
Phys. Lett. B \textbf{829} (2022), 137105
[arXiv:2001.01237 [hep-ph]].
\bibitem{Ellis:2022lft}
J.~Ellis, M.~Lewicki, M.~Merchand, J.~M.~No and M.~Zych,
JHEP \textbf{01}, 093 (2023)
doi:10.1007/JHEP01(2023)093
[arXiv:2210.16305 [hep-ph]].
\bibitem{Lewicki:2021pgr}
M.~Lewicki, M.~Merchand and M.~Zych,
JHEP \textbf{02}, 017 (2022)
doi:10.1007/JHEP02(2022)017
[arXiv:2111.02393 [astro-ph.CO]].
\bibitem{Cline:2020jre}
J.~M.~Cline and K.~Kainulainen,
Phys. Rev. D \textbf{101}, no.6, 063525 (2020)
doi:10.1103/PhysRevD.101.063525
[arXiv:2001.00568 [hep-ph]].
\bibitem{Carena:2025flp}
M.~Carena, A.~Ireland, T.~Ou and I.~R.~Wang,
[arXiv:2504.17841 [hep-ph]].
\bibitem{Croon:2023zay}
D.~Croon,
PoS \textbf{TASI2022}, 003 (2024)
doi:10.22323/1.439.0003
[arXiv:2307.00068 [hep-ph]].


\bibitem{Pati:1974yy}
J.~C.~Pati and A.~Salam,
Phys. Rev. D \textbf{10}, 275-289 (1974)
[erratum: Phys. Rev. D \textbf{11}, 703-703 (1975)]
\bibitem{Mohapatra:1974hk}
R.~N.~Mohapatra and J.~C.~Pati,
Phys. Rev. D \textbf{11}, 566-571 (1975)
\bibitem{Mohapatra:1974gc}
R.~N.~Mohapatra and J.~C.~Pati,
Phys. Rev. D \textbf{11}, 2558 (1975)
\bibitem{Senjanovic:1975rk}
G.~Senjanovic and R.~N.~Mohapatra,
Phys. Rev. D \textbf{12}, 1502 (1975)
\bibitem{Barenboim:1998ib}
G.~Barenboim and N.~Rius,
Phys. Rev. D \textbf{58}, 065010 (1998)
[arXiv:hep-ph/9803215 [hep-ph]].
\bibitem{Brdar:2019fur}
V.~Brdar, L.~Graf, A.~J.~Helmboldt and X.~J.~Xu,
JCAP \textbf{12}, 027 (2019)
[arXiv:1909.02018 [hep-ph]].
\bibitem{Li:2020eun}
M.~Li, Q.~S.~Yan, Y.~Zhang and Z.~Zhao,
JHEP \textbf{03}, 267 (2021)
[arXiv:2012.13686 [hep-ph]].
\bibitem{Wang:2024wcs}
D.~W.~Wang, Q.~S.~Yan and M.~Huang,
Phys. Rev. D \textbf{110}, no.7, 076011 (2024)
[arXiv:2405.01949 [gr-qc]].

\bibitem{Graf:2021xku}
L.~Gr\'af, S.~Jana, A.~Kaladharan and S.~Saad,
JCAP \textbf{05}, no.05, 003 (2022)
[arXiv:2112.12041 [hep-ph]].
\bibitem{Senjanovic:1978ev}
G.~Senjanovic,
Nucl. Phys. B \textbf{153}, 334-364 (1979)
\bibitem{Mohapatra:1977be}
R.~N.~Mohapatra and D.~P.~Sidhu,
Phys. Rev. D \textbf{16}, 2843 (1977)

\bibitem{Karmakar:2023ixo}
S.~Karmakar and D.~Ringe,
Phys. Rev. D \textbf{109}, no.7, 075034 (2024)
doi:10.1103/PhysRevD.109.075034
[arXiv:2309.12023 [hep-ph]].
\bibitem{Ma:1989tz}
E.~Ma,
Phys. Rev. Lett. \textbf{63}, 1042 (1989)
\bibitem{Borah:2017leo}
D.~Borah and A.~Dasgupta,
JCAP \textbf{06}, 003 (2017)
[arXiv:1702.02877 [hep-ph]].


\bibitem{Gu:2010yf}
P.~H.~Gu,
Phys. Rev. D \textbf{81}, 095002 (2010)
[arXiv:1001.1341 [hep-ph]].
\bibitem{Klimenko:1984qx}
K.~G.~Klimenko,
Theor. Math. Phys. \textbf{62}, 58-65 (1985)
\bibitem{Kannike:2012pe}
K.~Kannike,
Eur. Phys. J. C \textbf{72}, 2093 (2012)
[arXiv:1205.3781 [hep-ph]].
\bibitem{Boto:2022uwv}
R.~Boto, J.~C.~Rom\~ao and J.~P.~Silva,
Phys. Rev. D \textbf{106}, no.11, 115010 (2022)
[arXiv:2208.01068 [hep-ph]].
\bibitem{Langacker:1989xa}
P.~Langacker and S.~U.~Sankar,
Phys. Rev. D \textbf{40}, 1569-1585 (1989)
\bibitem{Workman:2022ynf}
R.~L.~Workman \textit{et al.} [Particle Data Group],
PTEP \textbf{2022}, 083C01 (2022)
\bibitem{Czakon:1999ga}
M.~Czakon, J.~Gluza and M.~Zralek,
Phys. Lett. B \textbf{458}, 355-360 (1999)
[arXiv:hep-ph/9904216 [hep-ph]].


\bibitem{Linde:1977mm}
A.~D.~Linde,
Phys. Lett. B \textbf{70}, 306-308 (1977)
doi:10.1016/0370-2693(77)90664-5
\bibitem{Linde:1981zj}
A.~D.~Linde,
Nucl. Phys. B \textbf{216}, 421 (1983)
[erratum: Nucl. Phys. B \textbf{223}, 544 (1983)]
doi:10.1016/0550-3213(83)90072-X
\bibitem{Coleman:1977th}
S.~R.~Coleman, V.~Glaser and A.~Martin,
Commun. Math. Phys. \textbf{58}, 211-221 (1978)
doi:10.1007/BF01609421
\bibitem{Espinosa:2018hue}
J.~R.~Espinosa,
JCAP \textbf{07}, 036 (2018)
doi:10.1088/1475-7516/2018/07/036
[arXiv:1805.03680 [hep-th]].
\bibitem{Croon:2020cgk}
D.~Croon, O.~Gould, P.~Schicho, T.~V.~I.~Tenkanen and G.~White,
JHEP \textbf{04}, 055 (2021)
doi:10.1007/JHEP04(2021)055
[arXiv:2009.10080 [hep-ph]].
\bibitem{Saikawa:2018rcs}
K.~Saikawa and S.~Shirai,
JCAP \textbf{05}, 035 (2018)
doi:10.1088/1475-7516/2018/05/035
[arXiv:1803.01038 [hep-ph]].
\bibitem{Ellis:2018mja}
J.~Ellis, M.~Lewicki and J.~M.~No,
JCAP \textbf{04}, 003 (2019)
doi:10.1088/1475-7516/2019/04/003
[arXiv:1809.08242 [hep-ph]].
\bibitem{Guth:1979bh}
A.~H.~Guth and S.~H.~H.~Tye,
Phys. Rev. Lett. \textbf{44}, 631 (1980)
[erratum: Phys. Rev. Lett. \textbf{44}, 963 (1980)]
doi:10.1103/PhysRevLett.44.631
\bibitem{Guth:1981uk}
A.~H.~Guth and E.~J.~Weinberg,
Phys. Rev. D \textbf{23}, 876 (1981)
doi:10.1103/PhysRevD.23.876
\bibitem{Guth:1982pn}
A.~H.~Guth and E.~J.~Weinberg,
Nucl. Phys. B \textbf{212}, 321-364 (1983)
doi:10.1016/0550-3213(83)90307-3
\bibitem{Enqvist:1991xw}
K.~Enqvist, J.~Ignatius, K.~Kajantie and K.~Rummukainen,
Phys. Rev. D \textbf{45}, 3415-3428 (1992)
doi:10.1103/PhysRevD.45.3415
\bibitem{Turner:1992tz}
M.~S.~Turner, E.~J.~Weinberg and L.~M.~Widrow,
Phys. Rev. D \textbf{46}, 2384-2403 (1992)
doi:10.1103/PhysRevD.46.2384
\bibitem{Espinosa:2008kw}
J.~R.~Espinosa, T.~Konstandin, J.~M.~No and M.~Quiros,
Phys. Rev. D \textbf{78}, 123528 (2008)
doi:10.1103/PhysRevD.78.123528
[arXiv:0809.3215 [hep-ph]].
\bibitem{Ellis:2020nnr}
J.~Ellis, M.~Lewicki and V.~Vaskonen,
JCAP \textbf{11}, 020 (2020)
doi:10.1088/1475-7516/2020/11/020
[arXiv:2007.15586 [astro-ph.CO]].
\bibitem{Baldes:2023rqv}
I.~Baldes and M.~O.~Olea-Romacho,
JHEP \textbf{01}, 133 (2024)
doi:10.1007/JHEP01(2024)133
[arXiv:2307.11639 [hep-ph]].
\bibitem{DiValentino:2021izs}
E.~Di Valentino, O.~Mena, S.~Pan, L.~Visinelli, W.~Yang, A.~Melchiorri, D.~F.~Mota, A.~G.~Riess and J.~Silk,
Class. Quant. Grav. \textbf{38}, no.15, 153001 (2021)
doi:10.1088/1361-6382/ac086d
[arXiv:2103.01183 [astro-ph.CO]].
\bibitem{Kosowsky:1991ua}
A.~Kosowsky, M.~S.~Turner and R.~Watkins,
Phys. Rev. D \textbf{45}, 4514-4535 (1992)
doi:10.1103/PhysRevD.45.4514
\bibitem{Kosowsky:1992rz}
A.~Kosowsky, M.~S.~Turner and R.~Watkins,
Phys. Rev. Lett. \textbf{69}, 2026-2029 (1992)
doi:10.1103/PhysRevLett.69.2026
\bibitem{Kosowsky:1992vn}
A.~Kosowsky and M.~S.~Turner,
Phys. Rev. D \textbf{47}, 4372-4391 (1993)
doi:10.1103/PhysRevD.47.4372
[arXiv:astro-ph/9211004 [astro-ph]].
\bibitem{Kamionkowski:1993fg}
M.~Kamionkowski, A.~Kosowsky and M.~S.~Turner,
Phys. Rev. D \textbf{49}, 2837-2851 (1994)
doi:10.1103/PhysRevD.49.2837
[arXiv:astro-ph/9310044 [astro-ph]].
\bibitem{Caprini:2007xq}
C.~Caprini, R.~Durrer and G.~Servant,
Phys. Rev. D \textbf{77}, 124015 (2008)
doi:10.1103/PhysRevD.77.124015
[arXiv:0711.2593 [astro-ph]].
\bibitem{Huber:2008hg}
S.~J.~Huber and T.~Konstandin,
JCAP \textbf{09}, 022 (2008)
doi:10.1088/1475-7516/2008/09/022
[arXiv:0806.1828 [hep-ph]].
\bibitem{Weir:2016tov}
D.~J.~Weir,
Phys. Rev. D \textbf{93}, no.12, 124037 (2016)
doi:10.1103/PhysRevD.93.124037
[arXiv:1604.08429 [astro-ph.CO]].
\bibitem{Jinno:2017fby}
R.~Jinno and M.~Takimoto,
JCAP \textbf{01}, 060 (2019)
doi:10.1088/1475-7516/2019/01/060
[arXiv:1707.03111 [hep-ph]].
\bibitem{Jinno:2019bxw}
R.~Jinno, T.~Konstandin and M.~Takimoto,
JCAP \textbf{09}, 035 (2019)
doi:10.1088/1475-7516/2019/09/035
[arXiv:1906.02588 [hep-ph]].
\bibitem{Lewicki:2020jiv}
M.~Lewicki and V.~Vaskonen,
Eur. Phys. J. C \textbf{80}, no.11, 1003 (2020)
doi:10.1140/epjc/s10052-020-08589-1
[arXiv:2007.04967 [astro-ph.CO]].
\bibitem{Megevand:2021juo}
A.~Megevand and F.~A.~Membiela,
JCAP \textbf{10}, 073 (2021)
doi:10.1088/1475-7516/2021/10/073
[arXiv:2108.05510 [astro-ph.CO]].
\bibitem{Giblin:2013kea}
J.~T.~Giblin, Jr. and J.~B.~Mertens,
JHEP \textbf{12}, 042 (2013)
doi:10.1007/JHEP12(2013)042
[arXiv:1310.2948 [hep-th]].
\bibitem{Giblin:2014qia}
J.~T.~Giblin and J.~B.~Mertens,
Phys. Rev. D \textbf{90}, no.2, 023532 (2014)
doi:10.1103/PhysRevD.90.023532
[arXiv:1405.4005 [astro-ph.CO]].
\bibitem{Hindmarsh:2015qta}
M.~Hindmarsh, S.~J.~Huber, K.~Rummukainen and D.~J.~Weir,
Phys. Rev. D \textbf{92}, no.12, 123009 (2015)
doi:10.1103/PhysRevD.92.123009
[arXiv:1504.03291 [astro-ph.CO]].
\bibitem{Hindmarsh:2016lnk}
M.~Hindmarsh,
Phys. Rev. Lett. \textbf{120}, no.7, 071301 (2018)
doi:10.1103/PhysRevLett.120.071301
[arXiv:1608.04735 [astro-ph.CO]].
\bibitem{Hindmarsh:2017gnf}
M.~Hindmarsh, S.~J.~Huber, K.~Rummukainen and D.~J.~Weir,
Phys. Rev. D \textbf{96}, no.10, 103520 (2017)
[erratum: Phys. Rev. D \textbf{101}, no.8, 089902 (2020)]
doi:10.1103/PhysRevD.96.103520
[arXiv:1704.05871 [astro-ph.CO]].
\bibitem{Hindmarsh:2019phv}
M.~Hindmarsh and M.~Hijazi,
JCAP \textbf{12}, 062 (2019)
doi:10.1088/1475-7516/2019/12/062
[arXiv:1909.10040 [astro-ph.CO]].
\bibitem{Caprini:2006jb}
C.~Caprini and R.~Durrer,
Phys. Rev. D \textbf{74}, 063521 (2006)
doi:10.1103/PhysRevD.74.063521
[arXiv:astro-ph/0603476 [astro-ph]].
\bibitem{Kahniashvili:2008pf}
T.~Kahniashvili, A.~Kosowsky, G.~Gogoberidze and Y.~Maravin,
Phys. Rev. D \textbf{78}, 043003 (2008)
doi:10.1103/PhysRevD.78.043003
[arXiv:0806.0293 [astro-ph]].
\bibitem{Kahniashvili:2008pe}
T.~Kahniashvili, L.~Campanelli, G.~Gogoberidze, Y.~Maravin and B.~Ratra,
Phys. Rev. D \textbf{78}, 123006 (2008)
[erratum: Phys. Rev. D \textbf{79}, 109901 (2009)]
doi:10.1103/PhysRevD.78.123006
[arXiv:0809.1899 [astro-ph]].
\bibitem{Kahniashvili:2009mf}
T.~Kahniashvili, L.~Kisslinger and T.~Stevens,
Phys. Rev. D \textbf{81}, 023004 (2010)
doi:10.1103/PhysRevD.81.023004
[arXiv:0905.0643 [astro-ph.CO]].
\bibitem{Caprini:2009yp}
C.~Caprini, R.~Durrer and G.~Servant,
JCAP \textbf{12}, 024 (2009)
doi:10.1088/1475-7516/2009/12/024
[arXiv:0909.0622 [astro-ph.CO]].
\bibitem{Kisslinger:2015hua}
L.~Kisslinger and T.~Kahniashvili,
Phys. Rev. D \textbf{92}, no.4, 043006 (2015)
doi:10.1103/PhysRevD.92.043006
[arXiv:1505.03680 [astro-ph.CO]].
\bibitem{RoperPol:2019wvy}
A.~Roper Pol, S.~Mandal, A.~Brandenburg, T.~Kahniashvili and A.~Kosowsky,
Phys. Rev. D \textbf{102}, no.8, 083512 (2020)
doi:10.1103/PhysRevD.102.083512
[arXiv:1903.08585 [astro-ph.CO]].
\bibitem{Athron:2022mmm}
P.~Athron, C.~Bal\'azs and L.~Morris,
JCAP \textbf{03}, 006 (2023)
doi:10.1088/1475-7516/2023/03/006
[arXiv:2212.07559 [hep-ph]].
\bibitem{Steinhardt:1981ct}
P.~J.~Steinhardt,
Phys. Rev. D \textbf{25}, 2074 (1982)
doi:10.1103/PhysRevD.25.2074
\bibitem{Espinosa:2010hh}
J.~R.~Espinosa, T.~Konstandin, J.~M.~No and G.~Servant,
JCAP \textbf{06}, 028 (2010)
doi:10.1088/1475-7516/2010/06/028
[arXiv:1004.4187 [hep-ph]].
\bibitem{Ellis:2019oqb}
J.~Ellis, M.~Lewicki, J.~M.~No and V.~Vaskonen,
JCAP \textbf{06}, 024 (2019)
doi:10.1088/1475-7516/2019/06/024
[arXiv:1903.09642 [hep-ph]].
\bibitem{Caprini:2015zlo}
C.~Caprini, M.~Hindmarsh, S.~Huber, T.~Konstandin, J.~Kozaczuk, G.~Nardini, J.~M.~No, A.~Petiteau, P.~Schwaller and G.~Servant, \textit{et al.}
JCAP \textbf{04}, 001 (2016)
doi:10.1088/1475-7516/2016/04/001
[arXiv:1512.06239 [astro-ph.CO]].
\bibitem{Schmitz:2020syl}
K.~Schmitz,
JHEP \textbf{01}, 097 (2021)
doi:10.1007/JHEP01(2021)097
[arXiv:2002.04615 [hep-ph]].
\bibitem{Davidson:2002qv}
S.~Davidson and A.~Ibarra,
Phys. Lett. B \textbf{535}, 25-32 (2002)
doi:10.1016/S0370-2693(02)01735-5
[arXiv:hep-ph/0202239 [hep-ph]].
\bibitem{Gu:2010zv}
P.~H.~Gu and M.~Lindner,
Phys. Lett. B \textbf{698}, 40-43 (2011)
doi:10.1016/j.physletb.2011.02.042
[arXiv:1010.4635 [hep-ph]].
\bibitem{Prokopec:2013ax}
T.~Prokopec, M.~G.~Schmidt and J.~Weenink,
Phys. Rev. D \textbf{87}, no.8, 083508 (2013)
doi:10.1103/PhysRevD.87.083508
[arXiv:1301.4132 [hep-th]].

\bibitem{Espinosa:2011eu}
J.~R.~Espinosa, B.~Gripaios, T.~Konstandin and F.~Riva,
JCAP \textbf{01}, 012 (2012)
doi:10.1088/1475-7516/2012/01/012
[arXiv:1110.2876 [hep-ph]].
\end{thebibliography}

\end{document}